\documentclass[12pt,column,A4paper]{article}

\usepackage[margin=1in]{geometry}
\usepackage{graphicx}
\usepackage{amsmath}
\usepackage{amssymb}
\usepackage{booktabs}
\usepackage{float}
\usepackage{indentfirst}

\usepackage[pagebackref,breaklinks,colorlinks]{hyperref}
\usepackage[nodisplayskipstretch]{setspace}
\newcommand\keywords[1]{\textbf{Keywords}: #1}

\usepackage[capitalize]{cleveref}
\crefname{section}{Sec.}{Secs.}
\Crefname{section}{Section}{Sections}
\Crefname{table}{Table}{Tables}
\crefname{table}{Tab.}{Tabs.}

\begin{document}

\title{Adjust factor with volatility model using MAXFLAT low-pass filter and construct portfolio in China A share market}

\author{Ke Zhang\\
{\tt\small kezzhang@gmail.com}
}
\date{}
\maketitle

%%%%%%%%% ABSTRACT
\begin{abstract}
   In the field of quantitative finance, volatility models, such as ARCH, GARCH, FIGARCH, SV, EWMA, play the key role in risk and portfolio management. Meanwhile, factor investing is more and more famous since mid of 20 century. CAPM, Fama–French three-factor model, Fama French five-factor model, MSCI Barra factor model are mentioned and developed during this period. In this paper, we will show why we need adjust group of factors by our MAXFLAT low-pass volatility model. All of our experiments are under China's CSI 300 and CSI 500 universe which represent China's large cap stocks and mid-small cap stocks. Our result shows adjust factors by MAXFLAT volatility model have better performance in both large cap and small cap universe than original factors or other risk adjust factors in China A share. Also the portfolio constructed by MAXFLAT risk adjust factors have continuous excess return and lower Beta compare with benchmark index. 

\end{abstract}
%%%%%%%%% BODY TEXT
\section{Introduction}

Since 1950s, factor model has playing more and more important role in finance field. Harry Markowitz proposed the Mean-Variance model\cite{markowitz1952portfilio} in 1952 which is the pioneer of modern finance theory. Then William Sharpe publishes the capital asset pricing model (CAPM)\cite{Sharpe1964capital} that firstly introduce the concept of factor, which explain the portfolio return by excess return and market exposure. In  1975, Barra mentioned the Barra factor models\cite{sheikh1996Barra}, and continuously developed it during last 50 years which help us forecast risk for equity, fixed income, cash and derivative instruments at portfolio and asset level. In 1993, Eugene Fama and Kenneth French presents the Famam-Fench three factor model\cite{fama1993common} which also include the bond factors. In 2015 Fama extend the factor model to five factors in his \cite{fama2015five}. Factor model are also widely used and researched in China market\cite{huang2019fama}\cite{cai2022research}\cite{fang2020case}\cite{wuestablishment}.

Risk model also play the key role in financial market, in the early of 1980s, one kind of famous volatility model was introduced by Engle, which is called Autoregressive conditionally heterosedastic model\cite{engle1982autoregressive}. Follow the work of Engle, in 1986, Bollerslev\cite{bollerslev1986generalized} introduced generalized autoregressive conditional heteroscedasticity model which allowed not only lagged in the square of errors but also historical variance. Meanwhile, other volatility model are also developed since then such as stochastic volatility and EWMA volatility model\cite{higgins1992class}. Later, people find in any stationary ARCH or GARCH model that memory decays exponentially fast. Therefore PWMA, IGARCH, FIGARCH\cite{baillie1996fractionally} or other long memory volatility were developed to deal with situation has long memory effect. For example, intra-day high frequency data. However, recently we can still find sometime people prefer to use rolling truncated period to estimate volatility which have better performance than all of volatility models mentioned above. After carefully research, we propose a new volatility with MAXFLAT loss-pass filter\cite{butterworth1930theory}\cite{andrew2010MAXFLAT} that the frequency response of this filter is flat in the pass-band (i.e. a bandpass filter) and roll-offs towards zero in the stop-band.. Also widely used rolling period realized volatility is one of special case of this volatility model. The advantage of this model is that volatility coefficients will decay slowly at beginning and fast at the end. In practice, we can use nonlinear state space representation of volatility model\cite{ke2016nonlinear}\cite{mullhaupt1997fast} with this MAXFLAT filter to estimate our volatility.

In this paper, base on the current research of factors model in China A share market, we firstly choose and construct several popular factors. Then we do some data pre-processing work to the stock universe and these factors. And we use our risk model to adjust these factors. We compare the performance of factors with different way to adjust. Finally we construct the portfolio with our risk adjust factors which outperform the benchmark in both large cap and mid-small cap universe.

\section{Data, universe and trade assumption}
 
Firstly, our model is low frequency that all of our data is captured at monthly base and we trade at each end of month. In order to watch the impact of our model to both large cap stocks and mid-small cap stocks, we pick two most popular universe in China A share market, CSI 300 and CSI 500. Our universe is dynamic and point-in time which avoid the survivor bias. 
On the other hand, we pick some famous and widely used factors\cite{zhu2017zhongmei}\cite{zhu2017lianghua} from open source ( https://uqer.datayes.com/). The factors generated from data until previous day of trading day to avoid look ahead. These factors can be Categorized as four group, value, profit, growth and tech which can represent multi-factors model that most researchers and analysts used in China A share market. 
We assume commission cost is 5 bps for buy and 1.5 bps for sell. Our order will totally be filled with 1 cent slippage cost for each side.
    
\begin{table}[H]
\centering
  \begin{tabular}{@{}lc@{}}
    \toprule
    Factor & Factor Category \\
    \midrule
    EP & Value \\
    EB & Value \\
    CFP & Value \\
    CTOP & Value\\
    CFO2EV & Quality\\
    ROE & Quality\\
    ROA & Quality\\
    GrossIncomeRatio & Quality\\
    ARTRate & Quality\\
    DebtsAssetRatio & Quality\\
    OperatingRevenueGrowRate & Growth\\
    NetProfitGrowRate & Growth\\
    SUE & Growth\\
    FEARNG & Growth\\
    FSALESG & Growth\\
    REVS20 & Tech\\
    VOL20 & Tech\\
    ILLIQUIDITY & Tech\\
    \bottomrule
  \end{tabular}
  \caption{Common factors we selected}
  \label{tab:t1}
\end{table}
\cref{tab:t1} shows the base style factors we get from open source https://uqer.datayes.com/.

\section{Data pre-processing}

Each end of month, we get stocks from our universe. However, we will not use all of these stocks to trade. We remove all the ST and ST* stocks, remove all the stocks be suspended or non-trade able, remove all the stocks listed less than 3 month, remove stocks whose 20 average day volume are less 10 million yuan. 

In order to calculate volatility, we need time series of underline asset. We take log of return series to make it closed to Gaussian distribution. Meanwhile we adjust the dividend and split for stock return.

As the priority of this paper is not factor analysis or find best factor. We just use simple rank to weight original factors. Lots of researchers has mentioned rank factor is a simple but useful and robust way to weight our portfolio. We firstly winsorize factor to remove outliers, then we rank each factor and get total score by average of each rank factor. As China A share market is hard to short, we want to build long only portfolio. Therefore rank factors automatically give us non-negative weight. 

\section{MAXFLAT low-pass filter}
MAXFLAT filter also known as Butterworth filter \cite{butterworth1930theory} which is firstly mentioned by Butterworth at 1930. The MAXFLAT low-pass filter is a type of signal processing filter designed to have a frequency response that is as flat as possible in the pass-band and rolls off towards zero in the stop-band. As we want our volatility Coefficient decay slowly at beginning and fast at end, MAXFLAT filter easily satisfied our requirement. Also, widely used rolling window realized volatility is one of spacial case of MAXFLAT filter.
    
Here is some good property of MAXFLAT low-pass filter.
\begin{itemize}
\item Maximal flatness in the pass-band and zero in the stop-band.
\item Monotonically decrease from the specified cut-off frequencies.
\item Smooth response at all frequencies.
\item Half-power frequency that corresponds to the specified cut-off frequencies.
\end{itemize}

Then let's see the detail and formula of MAXFLAT low-pass filter.

\begin{equation}
  |H_n(jw)| = \frac{1}{\sqrt{1+w^{2n}}}
\label{eq:e1}
\end{equation}
\cref{eq:e1} is MAXFLAT low-pass frequency response. 

Where w is the critical frequency or frequencies and n is the order of the filter. 

Take square and substitute s = jw in \cref{eq:e1}, we get 
\begin{equation}
  H(S)H(-S) = \frac{1}{1+{(s/j)}^{2n}}
\label{eq:e5}
\end{equation}

The poles of $H(S)H(-S)$ are given by:

\begin{equation}
  1+(s/j)^{2n} = 0 \xrightarrow{}\\
  s^{2n} = -1j^{2n}
\label{eq:e6}
\end{equation}

As 

\begin{equation}
  -1 = e^{j\pi{(2k-1)}},j = e^{j\pi/2}
\label{eq:e4}
\end{equation}

We can derive:

\begin{equation}
  s^{2n} = e^{j\pi(2k-1+n)}
\label{eq:e7}
\end{equation}

Therefore the poles of $H(S)H(-S)$ line on the unity circle:

\begin{equation}
  s_k = e^{\frac{j\pi(2k-1+n)}{2n}}\\
  , k = 1,2,3,...,2n
\label{eq:e8}
\end{equation}

As we are low-pass filter, we only care $H(S)$ which have poles line on the left side of unity circle. So the poles of our MAXFLAT low-pass filter are:

\begin{equation}
  s_k = e^{\frac{j\pi(2k-1+n)}{2n}}\\
  , k = 1,2,3,...,n
\label{eq:e9}
\end{equation}

Then the transfer function of our low-pass filter is:

\begin{equation}
  H(s) = \frac{1}{\prod_i(s-s_i)}\\
  , i = 1,2,3,...,n
\label{eq:e10}
\end{equation}

\section{MAXFLAT low-pass non-linear state space volatility representation}

As we already have numerator b and denominator a, we can also get state space representation in controller canonical form of our filter. As volatility related to square of return, we need to consider the non-linear state space representation. To simplify, we use inverse of system representation here.
\begin{equation}
\begin{split}
  Z(t+1) = AZ(t) + By^2(t) \\
  x^2(t) = y^2(t)/CZ(t)
\label{eq:e13}
\end{split}
\end{equation}
\cref{eq:e13} is our non-linear State space representation of volatility\cite{ke2016nonlinear}.
Here y(t) is demean log return at time t (If we didn't demean y(t) here, there will be a constant term in state space representation). C,A,B is parameters of state space model get from MAXFLAT transfer function. x(t) is white noise. 

\begin{equation}
\begin{aligned}
  \sigma^2(t) = \sum_i{CA^iB y^2(t-i)}\\
  st. CA^kB>0,\sum_k{CA^kB}<1
\label{eq:e14}
\end{aligned}
\end{equation}
Then we can derive square of volatility at time t though \cref{eq:e14}

\section{Volatility model with different decay}

\begin{figure}[H]
\centering
  %\fbox{\rule{0pt}{2in} \rule{0.9\linewidth}{0pt}}
  \includegraphics[width=.8\linewidth]{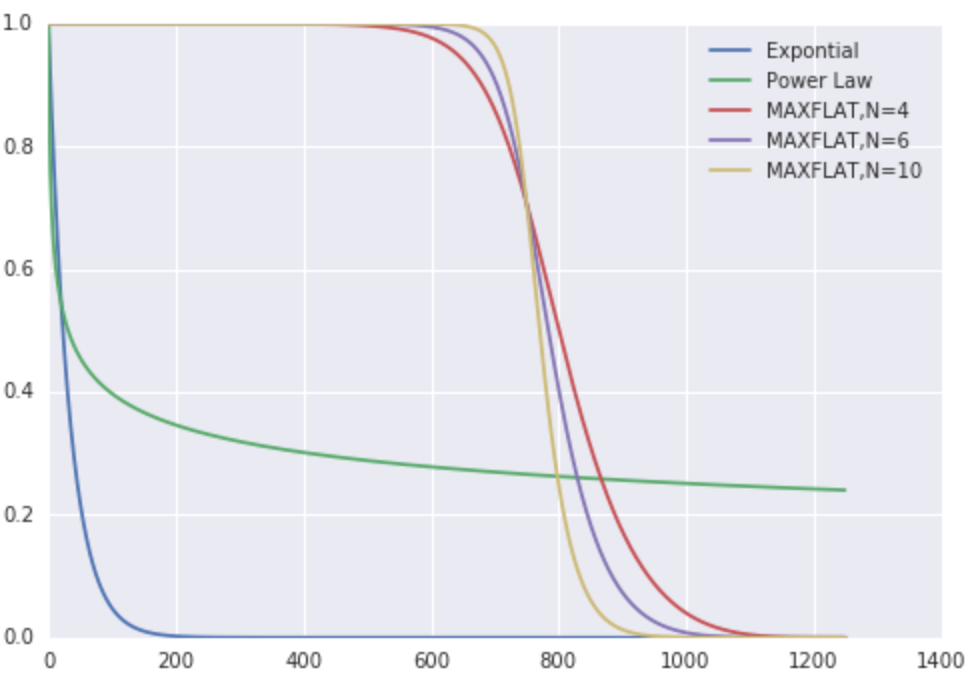}
   \caption{Decay of different methods}
  \label{fig:b_decay}
\end{figure}
The above figures shows how fast each methods decay. PWMA decay fastest at beginning and slowest at the end. EWMA decay faster at beginning than MAXFLAT and fastest finally. MAXFLAT decay very slow firstly and faster from the middle.

\begin{table}[H]
\centering
  \begin{tabular}{@{}lc@{}}
    \toprule
    Coefficient Decay & Volatility Model \\
    \midrule
    Exponential & EWMA,ARCH,GARCH,SV \\
    Power law & PWMA,IGARCH,FIGARCH \\
    MAXFLAT & Nonlin State Space Vol\\ 
    
    \bottomrule
  \end{tabular}
  \caption{Vol Model by decay speed}
  \label{tab:t2}
\end{table}
\cref{tab:t2} show the category of some popular volatility model by decay speed. Recently most of long memory Volatility model are decay similar with power law speed. Our volatility model with MAXFLAT low-pass filter is a innovation in volatility field.

\section{Model construction and comparison}
After we pre processed our data, then we consider to construct our portfolio. Lots of researchers consider volatility as one of factors and linearly add volatility factors with other factors. We have different opinion that we linearly weight all Value, Quality, Growth and Tech factors then we divide the combined factors by our volatility estimator.
For volatility model, for comparison, we pick some famous candidates. Firstly, we choose EWMA who also known as risk-metric used firstly by JP Morgan. EWMA is also spacial case of GARCH(1,1). Volatility of EWMA or GARCH model decay exponentially which is too fast so people consider some long memory volatility model such as PWMA, FIGARCH\cite{baillie1996fractionally}. These model mostly decay with power law. In practice, we can also find lots of researchers or analyst using fix rolling window to estimator realized volatility. Here we use 1 year and 2 year realized volatility as our candidates. Finally, we construct a portfolio with adjust factors by MAXFLAT low-pass filter. 

\begin{table}[H]
\centering
\begin{tabular}{||c c c||} 
 \hline
 Volatility Methods & Total Return (\%) & Sharpe Ratio  \\ [1ex] 
 \hline\hline
 EWMA & 120.69 &  0.48  \\ 
 \hline
 250VOL & 183.48 & 0.61 \\
 \hline
 500VOL & 219.04 & 0.68  \\
 \hline
 PWMA & 243.09 & 0.72 \\
 \hline
 NoVOL & 149.46 & 0.44  \\ 
\hline
 MAXFLAT & 242.86 & 0.72 \\  [1ex] 
 \hline
\end{tabular}
\caption{Different Volatility Estimator with CSI 300 universe.}
\label{tab:t3}
\end{table}

In \cref{tab:t3}, we construct portfolio with adjust factor by different Volatility model in universe CSI 300. For each portfolio, we pick first 50 stocks by rank of adjust factor score. For NoVOL(no Volatility adjustment), it has the worst Sharpe ratio and second worst total return which tell us we definitely need do risk adjustment. On the other hand, EWMA result better than NoVol but worst than other model. Which means, focus on short term volatility is a bad idea for factor adjustment. MAXFLAT and PWMA has best total return and Sharpe ratio among all group in this universe. 

\begin{table}[H]
\centering
\begin{tabular}{||c c c||} 
 \hline
 Volatility Methods & Total Return (\%) & Sharpe Ratio  \\ [1ex] 
 \hline\hline
 EWMA & 144.13 & 0.45  \\ 
 \hline
 250VOL & 202.17 & 0.52 \\
 \hline
 500VOL & 206.71 & 0.53  \\
 \hline
 PWMA & 185.49 & 0.50  \\
 \hline
 NoVOL & 151.46 & 0.41  \\ 
\hline
 MAXFLAT & 208.51 & 0.53  \\  [1ex] 
 \hline
\end{tabular}
\caption{Different Volatility Estimator with CSI 500 universe.}
\label{tab:t4}
\end{table}

In \cref{tab:t4}, here we construct portfolio with adjust factor by different volatility model in universe CSI 500. For each portfolio, we pick first 50 stocks by rank of adjust factor score. NoVOL(no Volatility adjustment) and EWMA are still the worst group. Which means, No Volatility adjustment or just focus on short term volatility is a bad idea for factor adjustment in CSI 500. MAXFLAT has best total return and Sharpe ratio in all group in CSI 500 universe.

No matter in CSI 300 or CSI 500 universe, MAXFLAT Volatility adjustment show the best result in term of return and Sharpe ratio. We guess the reason are volatility clustering effect and long term instead of short term low volatility of stocks really improve return and Sharpe ratio of portfolio. 

\section{Split MAXFLAT factor portfolio by quantile}

In this section, we first construct factor portfolio by rank of MAXFLAT adjust factor score. Then we split the portfolio to several group by quantile. We want to see if higher adjust factor score monotonically improve performance of portfolio.

\begin{figure}[H]
\centering
  \includegraphics[width=.8\linewidth]{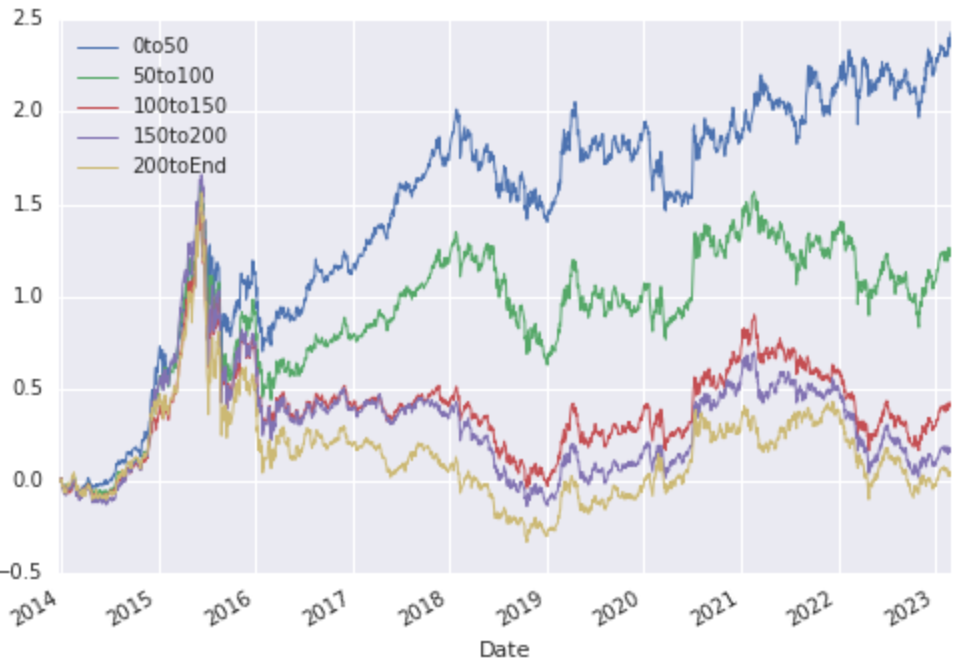}
   \caption{P$\&$L Quantile with CSI 300 universe using MAXFLAT low-pass volatility model}
   \label{fig:b_quantile300}
   
\end{figure}

In \cref{fig:b_quantile300}, we construct MAXFLAT factor portfolio in CSI 300. We split the portfolio to several group by quantile of adjust factor score. In this universe, we can find group with higher factor score has better total return monotonically.

\begin{figure}[H]
\centering
  \includegraphics[width=.8\linewidth]{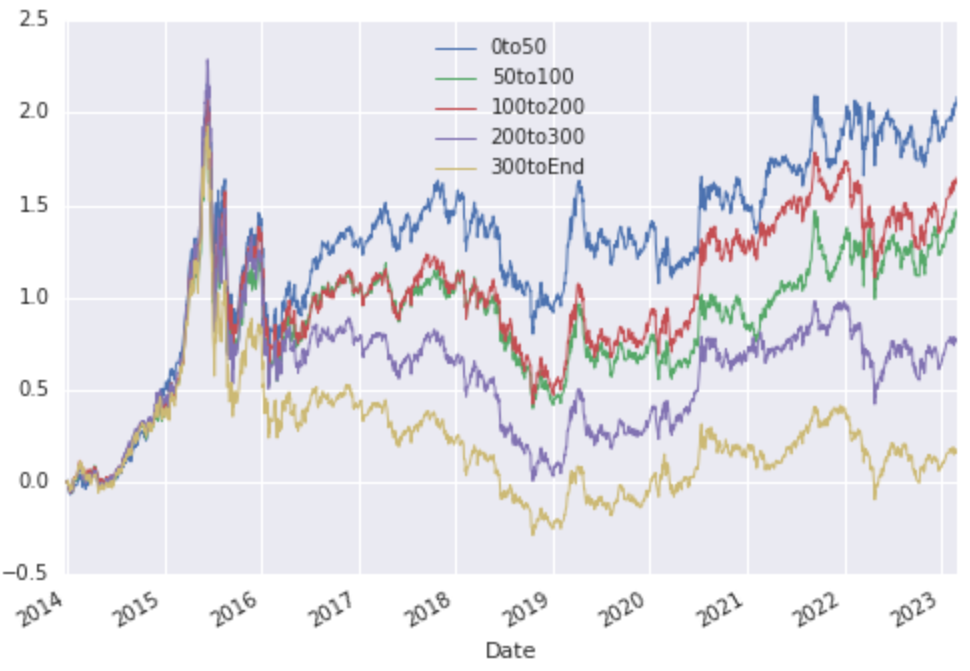}
   \caption{P$\&$L Quantile with CSI 500 universe using MAXFLAT volatility model}
   \label{fig:b_quantile500}
\end{figure}

In \cref{fig:b_quantile500}, we construct MAXFLAT factor portfolio in CSI 500 universe. In this universe, except 50 to 100 is slightly worse than 100 to 200, all other group with higher factor score monotonically has better total return.

As the result above, we showed higher factor score monotonically has better total return in both large cap and mid-small cap universe in China A share market.

\subsection{Compare MAXFLAT factor portfolio with benchmark index}

In this section, for CSI 300 and CSI 500 universe, we construct MAXFLAT factor portfolio with first 50 stocks that have highest MAXFLAT adjust factor score. Then we compare portfolio performance with corresponding index. 

\begin{figure}[H]
\centering
  \includegraphics[width=.8\linewidth]{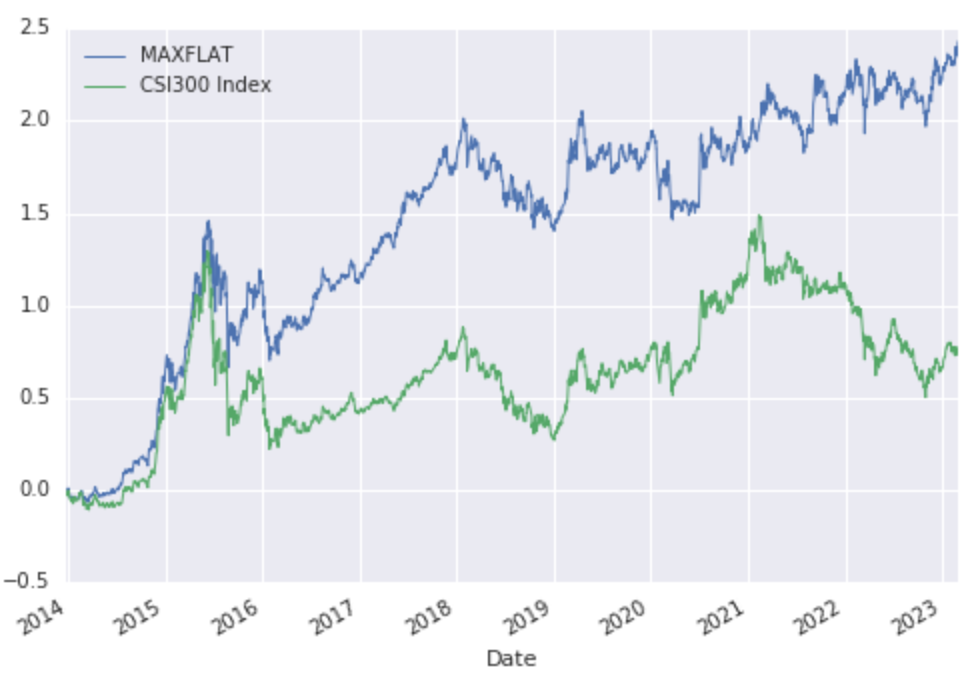}
   \caption{P$\&$L compare with benchmark with CSI 300 universe using MAXFLAT volatility model}
   \label{fig:b_bench300}
\end{figure}

\begin{table}[H]
\centering
\begin{tabular}{||c c c c c||} 
 \hline
 Porfolio & Total Return (\%) & Sharpe & Alpha & Beta   \\[0.8ex] 
 \hline\hline
 MAXFLAT & 242.86 & 0.72 & 8.84 & 0.75 \\ 
 \hline
 CSI 300 Index & 76.94 & 0.29 & 0 & 1 \\
 \hline
\end{tabular}
\caption{Different Volatility Estimator wtih CSI 300 universe.}
\label{tab:t5}
\end{table}

In \cref{tab:t5}, We construct MAXFLAT porfolio and compare with benchmark in CSI 300 universe. MAXFLAT portfolio beat index in term of total return, Sharpe ratio, Alpha and Beta.

\begin{figure}[H]
\centering
  \includegraphics[width=.8\linewidth]{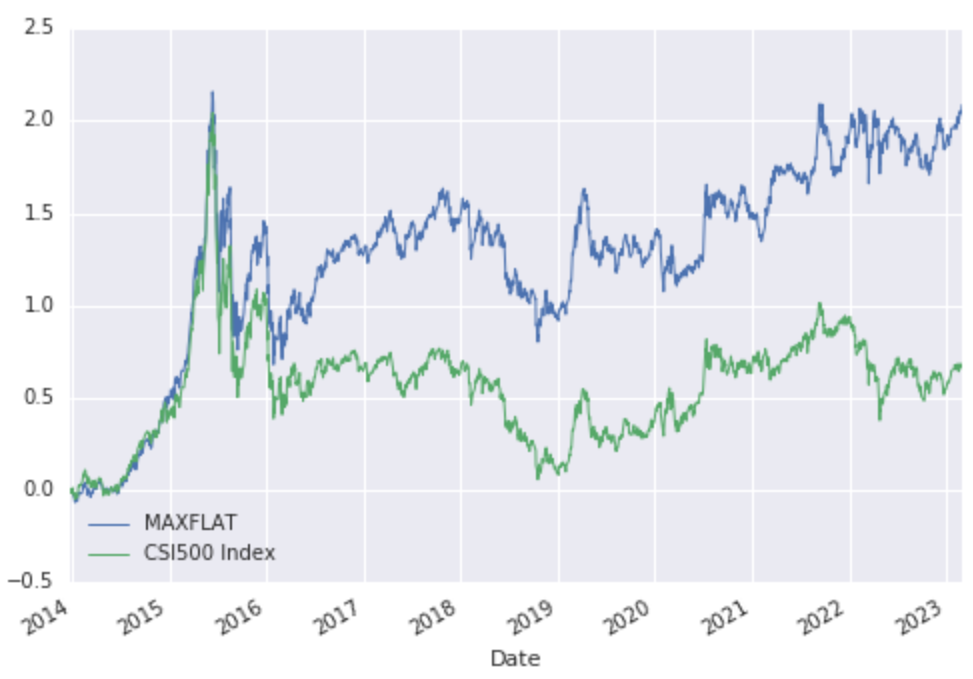}
   \caption{P$\&$L compare with benchmark with CSI 500 universe using MAXFLAT low-pass volatility model}
   \label{fig:b_bench500}
\end{figure}

\begin{table}[H]
\centering
\begin{tabular}{||c c c c c||} 
 \hline
 Porfolio & Total Return (\%) & Sharpe & Alpha & Beta   \\ [0.8ex] 
 \hline\hline
 MAXFLAT & 208.51 & 0.53 & 7.66 & 0.83 \\ 
 \hline
 CSI 500 Index & 68.70 & 0.23 & 0 & 1 \\
 \hline
\end{tabular}
\caption{Different Volatility Estimator wtih CSI 500 universe.}
\label{tab:t6}
\end{table}
In \cref{tab:t6}, Here we compare MAXFLAT portfolio with benchmark in CSI 500 universe. MAXFLAT portfolio beat this index for total return, Sharpe ratio, Alpha and Beta.

\section{Conclusion}
In this paper, we definite a new volatility model by MAXFLAT low-pass filter. We construct the adjust multi-factors model with different volatility model for comparison. And we show our MAXFLAT low-pass volatility model outperform other volatility adjustment in China A share market CSI 300 and CSI 500. The higher adjust factor score the quantile of portfolio have better performance. Meanwhile, no matter in universe CSI 300 or CSI 500, our model beat corresponding benchmark index. The result explain the volatility clustering effect and the importance of stock long term volatility for portfolio selection.
%%%%%%%%% 

\end{document}